# Impedance spectroscopy of epitaxial multiferroic thin films

[1]Rainer Schmidt*, [1]Wilma Eerenstein, [2]Thomas Winiecki, [3]Finlay D. Morrison, [1]Paul A. Midgley

[1]*University of Cambridge, Department of Materials Science, Pembroke Street, Cambridge CB2 3QZ, UK*

[2] *University of Durham, Department of Physics, South Road, Durham DH1 3LE, UK*

[3]*University of Cambridge, Department of Earth Sciences, Downing Street, Cambridge CB2 3EQ, UK*

* corresponding author: rainerxschmidt@googlemail.com

Temperature dependent impedance spectroscopy enables the many contributions to the dielectric and resistive properties of condensed matter to be deconvoluted and characterized separately. We have achieved this for multiferroic epitaxial thin films of $BiFeO_3$ (BFO) and $BiMnO_3$ (BMO), key examples of materials with strong magneto-electric coupling. We demonstrate that the true film capacitance of the epitaxial layers is similar to that of the electrode interface, making analysis of capacitance as a function of film thickness necessary to achieve deconvolution. We modeled non-Debye impedance response using Gaussian distributions of relaxation times and reveal that conventional resistivity measurements on multiferroic layers may be dominated by interface effects. Thermally activated charge transport models yielded activation energies of 0.60 eV ± 0.05 eV (BFO) and 0.25 eV ± 0.03 eV (BMO), which is consistent with conduction dominated by oxygen vacancies (BFO) and electron hopping (BMO). The intrinsic film dielectric constants were determined to be 320 ± 75 (BFO) and 450± 100 (BMO).

# I. INTRODUCTION

Multiferroic materials have attracted renewed interest in recent years.[1-4] $BiFeO_3$ (BFO) is one of the most extensively studied multiferroics, because it is the only material known to exhibit magnetic and ferroelectric order at room temperature.[5] $BiMnO_3$ (BMO) has attracted great interest due to a large magnetic moment of 3.6 $\mu_B$/Fe below $T_C^{FM}$ = 105 K in polycrystals despite being electrically insulating.[6] Magnetoelectric coupling in BFO films [7,8] and BMO polycrystals [9] has been claimed, implying that an applied magnetic field can induce changes in the ferroelectric order and an electric field changes to the magnetization. Coupling of magnetic and electric order parameters is of great technological and fundamental importance, and can be studied directly by measuring changes in polarization induced by an applied magnetic field, or by measuring changes in magnetization by an applied electric field.[2] These measurements are affected though by current leakage, which prevents the application of sustained electric fields and, in the case of polarization measurements, contributes to the measured current. An alternative way to investigate magnetoelectric coupling is the measurement of the magneto-capacitance.[9] Impedance spectroscopy (IS) is then of great interest as a technique to reliably determine the intrinsic film dielectric constant to enable accurate magneto-capacitance measurements. Unwanted contributions such as those at the electrode interface can be deconvoluted, and leakage currents can be accounted for by modeling the intrinsic film impedance response by a parallel resistor-capacitor circuit (RC element), where the parallel resistor describes the leakage behavior of the ferroelectric capacitor.

# II. IMPEDANCE SPECTROSCOPY (IS)

IS involves the application of an alternating voltage signal to a sample, and the measurement of the phase shifted current response. Fig. 1 presents the basic principle of an IS experiment and the definitions of the complex impedance of various circuit elements. It has been shown previously that different dielectric relaxation processes detected by IS in a polycrystal such as those found at the electrode interface, grain boundary and the intrinsic bulk regions can, in the simplest case, all be described by an RC circuit element consisting of a resistor and capacitor in parallel.[10,11] The macroscopic impedance is then just the sum of all series RC impedances. The magnitude of each specific capacitance $c$ (= $\varepsilon_0 \varepsilon'$) [F cm$^{-1}$] may identify the origin of the relaxation,[11] where $\varepsilon_0$ and $\varepsilon'$ are the permittivity of vacuum and the relative permittivity of a specific contribution. In this paper we show that IS and the $c$ classification scheme can be applied to multiferroic epitaxial layers, and the film and interface impedance can be modeled using series RC elements.

In reality, few systems can be represented by ideal resistors and capacitors. In order to account for non-Debye behavior, a well established approach is the replacement of $c$ by a phenomenological Constant-Phase-Element (CPE),[10] and indeed was successfully applied in this study. Such a parallel ideal resistor and non-ideal CPE capacitance circuit (R-CPE) has a specific complex impedance of

$$z^*_{R-CPE} = \frac{\rho_{dc}}{1 + \rho_{dc} c_m (i\omega)^n} \quad (1)$$

where $\omega$ is the angular frequency of the time ($t$) dependent applied alternating voltage signal. To account for the empirical parameter $n$, the specific capacitance $c$ is modified ($c_m$) and has units of F s$^{n-1}$ cm$^{-1}$. $\rho_{dc}$ is the D.C. resistivity of the resistor in $\Omega$ cm. The origin of non-Debye response is difficult to determine and several possibilities have been proposed.[10] We suggest that in stable epitaxial films the use of a random distribution $\Delta\tau$ of relaxation times $\tau$ (= $\rho_{dc} c$) may be the physically most meaningful approach to model non-Debye impedance behavior.[12] It is plausible that $\tau$ always shows a certain distribution across a sample. In fact, the constituent parameters of $\tau$, i.e. $\rho_{dc}$ and $c$, may both display independent distributions; and further, microscopically, both parameters may again depend on several material specific parameters with independent distributions each. To account for such complexity a Gaussian distribution of $\tau$ may be the best approximation possible. This approximation is based on the fundamental principle of the "central limit theorem", which states that functions of a large number of independent or weakly-dependent random variables have a probability distribution close to the normal distribution.[13] This is supported experimentally by a recent study of microscopic single grain boundary relaxation times in SrTiO$_3$ (STO), which show a Gaussian type distribution.[14] For electrode interface effects, non-Debye behavior has been associated with the fractal nature of the interface with dimensionality $d$ of $2 < d < 3$, where $d$ can be directly related to $n$.[15] RC elements with a random Gaussian distribution of relaxation times have an impedance of [12]:

$$z^*_{Voigt} = \frac{\rho_{dc}}{\Delta\tau\ \sqrt{2\pi}} \int_{-\infty}^{+\infty} \frac{\exp\left(\frac{-x^2}{2\ \Delta\tau^2}\right)}{1+i\omega\ (\bar{\tau}+x)}\ dx \qquad (2)$$

where $\bar{\tau}$ is the mean relaxation time, and $\Delta\tau$ the standard deviation of the Gaussian distribution. We have solved the integral in eq. (2) analytically and confirmed the solution numerically. A MatLab® program was developed to fit experimental data by a least mean squares fitting routine. The impedance of an ideal RC element is a Lorentzian function [eq. (1) with $n = 1$], and the convolution of Lorentzian and Gaussian functions is a Voigt profile. Such Voigt type profiles are commonly used in physical measurements, for example to describe the Lorentzian and Gaussian broadening of atomic or molecular spectral lines.[16] Here, Voigt profiles are used to describe impedance response of solid matter in order to replace the purely phenomenological CPE. Additionally, we developed fitting software based on the CPE circuit in eq. (1).

Impedance spectroscopy data can be plotted as negative imaginary vs the real part of the specific impedance $-z''$ vs $z'$, where for $n = 1$ each RC element is represented by a semicircle of radius $\rho_{dc}/2$ and maximum at $\omega_{max} = \tau^{-1}$. R-CPE or Voigt circuits yield a suppressed semicircle. From the complex specific impedance $z^*$ (= $z'- iz''$) we determined the BFO and BMO thin film complex dielectric constant $\varepsilon^*$ (= $\varepsilon'- i\varepsilon''$) and specific capacitance $c^*$ (= $c'-ic''$) from the complex relationship $z^* = (i\omega \varepsilon_0 \varepsilon^*)^{-1}$, where $\varepsilon^* = c^*/\varepsilon_0$.[10]

## III. THE BiFeO$_3$ (BFO) AND BiMnO$_3$ (BMO) SYSTEMS

BFO and BMO have previously been grown and studied in the form of single crystals, bulk ceramics and thin films, but comparatively low resistivity (i.e. high leakage) in such materials has hampered the study of ferroelectric polarization and magnetoelectric coupling. This is reflected in reports of a large range of saturation polarization values $P_s$: for BFO 2.2 - 158 µC cm$^{-2}$.[17-21] Such problems can be minimized in thin highly crystalline epitaxial layers with low leakage in the absence of extended grain boundaries and defects. However, epitaxial layers often show distinctively different physical properties compared to bulk due to epitaxial constraint. In the following we briefly review the properties of bulk and epitaxial BFO and BMO systems.

### A. BiFeO$_3$ (BFO)

Single crystal BFO has a rhombohedrally distorted perovskite structure ($a = b = c = 5.63$ Å; $\alpha = \beta = \gamma = 59.3°$),[22,23] and is an antiferromagnet with $T_N = 645$ K.[24,25] Fe$^{3+}$ cations are coupled anti-ferromagnetically, where the magnetic moments are canted and form a spiral spin structure with a wavelength of ~ 62 nm resulting in a zero net magnetic moment.[26] In epitaxial BFO grown on SrTiO$_3$ (STO) the latent magnetization can be released resulting in magnetic moments of ~ 0.02 - 0.05 µ$_B$/Fe.[27-29] This may be associated to the absence of spiral spin structures due to epitaxial constraint [29] and/or the fact that the film thickness is comparable to the spiral wavelength. A larger magnetic moment of ~ 0.5 $\mu_B$/Fe in epitaxial BFO has been reported,[7] which was associated with incomplete oxygenation and mixed $Fe^{2+}/Fe^{3+}$ valence in $BiFeO_{3-x}$.[30] In our epitaxial films the unit cell was indexed as pseudo-cubic with a small tetragonal elongation of $c/a = 1.027$ with $a = b = 3.905$ Å (identical to the STO substrate). Previously, a small monoclinic distortion of ~ 0.5° has been claimed as well.[7]

Ferroelectricity arises from the displacement of $Fe^{3+}$ cations from the centro-symmetric positions (by ~ 0.134 Å in bulk) in the $FeO_6^{9-}$ perovskite octahedra and the resulting dipole moment. In single crystals, the transition temperature $T_C^{FE}$ is ~ 1100 K [31] and a spontaneous polarization of 3.5 µC cm$^{-2}$ was reported.[5] In polycrystalline bulk ceramics an 8.9 µC cm$^{-2}$ saturation polarization was reported.[18] In BFO films a wide range of different room temperature remnant polarizations $2P_r$ have been found: 1.7 - 136 µC cm$^{-2}$ in granular films [17, 32-34] and 0.9 - 300 µC cm$^{-2}$ in epitaxial layers. [7,19-21,29,30,35] The coercive field in epitaxial layers was reported to be ~ 200 kV cm$^{-1}$.[7]

Magnetoelectric coupling in BFO films at room temperature has been claimed [7,8], an effect which has been observed at low temperatures in nickel iodine boracite,[36] and in the orthorhombic manganites $TbMnO_3$,[37] and $TbMn_2O_5$,[38] all exhibiting magnetic ordering transitions below room temperature. BFO is thus a model magnetoelectric material for studies at room temperature and above. The BFO dielectric constant in granular films has been reported to be ~ 110 [17] and ~ 140,[34] and ~ 80 for polycrystalline bulk samples.[39,40]

**B. BiMnO₃**

BiMnO₃ (BMO) has attracted interest due to coexisting ferroelectricity and ferromagnetism at low temperature with a large magnetic moment [6] despite the material being an insulator. Below the bulk transition temperature $T_C^{FM}$ ~ 105 K [6,41,42] a magnetic moment of ~ 3.6 $\mu_B$/Fe [6] has been reported. Single crystal BMO has a triclinically distorted perovskite structure ($a = c = 3.935$ Å; $\alpha = \gamma = 91.4°$; $b =$

3.989; $\beta = 91.0°$),[43] which can be represented by a monoclinic unit cell.[44] Ferromagnetism may be attributed to orbital ordering that produces three-dimensional ferromagnetic super-exchange interaction of $e_g$ electrons.[9] Ferroelectricity again arises from the displacement of $Mn^{3+}$ cations from the centro-symmetric positions in the $MnO_6^{9-}$ perovskite octahedra and the resulting dipole moment. The stabilization of the ferromagnetism and the ferroelectric off-centre distortion has been associated with the presence of Bi 6s lone pairs.[45] Various temperatures have been reported for the polycrystalline bulk ferroelectric transition temperature: $T_C^{FE} \sim 450$ K,[46] 500 K [41] and 750 K - 770 K.[9] The remnant bulk polarization $2P_r$ was claimed to be 86 nC cm$^{-2}$ at 200K.[46] In epitaxial films a ferromagnetic moment of 2.2 $\mu_B$/Fe below $T_C^{FM} = 85$ K has been reported.[44] The film remnant polarization $2P_r$ for granular films was claimed to be $\sim 8.2$ nC cm$^{-2}$ with a coercive field of $\sim 160$ kV cm$^{-1}$ and $T_C^{FE}$ of $\sim 450$ K.[46] In epitaxial layers on STO, the unit cell can be indexed as pseudo-cubic with $a = b = 3.905$ Å (identical to the STO substrate). The tetragonal $c$ elongation has been reported to be $c/a = 1.015$.[47] A possible monoclinic distortion is not known. Magnetoelectric coupling in polycrystals has been claimed from magnetocapacitance measurements and the dielectric constant was found to be temperature dependent and $\sim 28$ at 150 K.[9]

## IV. EXPERIMENTAL

**A. Epitaxial growth by Pulsed Laser Deposition (PLD)**

Epitaxial BFO and BMO films were grown on 1 at.% Nb doped STO (001) using Pulsed Laser Deposition (PLD) in an optimized procedure as described previously, including characterization by X-ray diffraction.[27,44] Uniform strain was confirmed in films of thickness $\leq$ 100nm and narrow (002) FWHM rocking curves of 0.05° (BFO) and 0.04° (BMO) (compared to 0.03° for STO substrates) were found. Film thickness was determined from X-ray fringes in $\omega$-$2\theta$ scans with an uncertainty $\leq$ 10%. Pt top electrodes were sputter-deposited through a mechanical mask.

**B. Impedance spectroscopy measurements**

The low resistivity of the Nb-STO substrate (~ 5 mΩ cm) [48] allowed the impedance to be measured across the film normal axis as indicated by the current path in Fig. 2. In ferroelectric thin films single domain structure along the film normal axis can be assumed,[49] which implies that A.C. currents flow within a single ferroelectric domain. Magnetic domains are absent in BMO above $T_C^{FM}$ and may be large in the case of the anti-ferromagnetic domains found in BFO implying single magnetic domain structure. Sample contacts were spring loaded stainless steel probes connected to Cu wires, which were interfaced with coaxial cables. IS was carried out at frequencies of 40 Hz – 2 MHz using an Agilent 4294A Impedance Analyzer with a signal amplitude of 50 mV, resulting in electric fields smaller than all reported coercive fields in BFO and BMO by at least a factor of 40. The sample temperature was varied between 25ºC – 300ºC using a custom-built temperature-controlled furnace.

**C. Data Analysis by equivalent circuit fitting**

The impedance spectra obtained have been fitted first using two conventional R-CPE elements for interface and film contributions (eq.1). Alternative models including ideal RC elements, conventional Debye elements of one ideal resistor and two capacitors accounting for high and low frequency limiting dielectric constants $\varepsilon(0)$ and $\varepsilon(\infty)$,[10] CPE-CPEs, and combinations of R-CPE and C-CPE elements, all clearly showed larger fitting errors, which were obtained from fitting software. All circuits were extended by a single resistor (R0) describing the resistance of the measurement probes, leads, Pt electrodes and Nb-STO substrate, and a residual inductance (L0) describing the inductance of the leads (Fig. 2). The residual parallel capacitance C0 describing the probe holder shown in Fig. 2 was neglected. C0 is at least 5 orders of magnitude lower than the film capacitance due to a comparatively large probe distance. The probe holder material PTFE (Poly-Tetra-Fluoro-Ethylene) has a dielectric constant of ~ 2, i.e. a factor of ~ 200 smaller than the films (determination of the film dielectric constants is described in detail below). The resistivity of PTFE is ~ $10^{18}$ Ω·cm and the parallel resistance can be neglected as well.

Fits to the data were obtained for 50 nm, 100 nm and 200 nm films at room temperature. Capacitance values obtained from CPEs in F $s^{n-1}$ $cm^{-1}$ were corrected to F $cm^{-1}$ using a standard method.[50]

## V. RESULTS AND DISCUSSIONS

### A. Deconvolution of interface and intrinsic film contribution

C1 and C2 values obtained from the fits were of the same order of magnitude, and were both normalized to the contact area $A$ and geometrical factor $g$ (= $A$ / [2 x film thickness]). Fig. 3 demonstrates that for BFO films normalization by $A$ leads to approximately constant C1 as expected for an interface contribution resulting in capacitance values typical of an interface, and C2 is approximately constant for normalization by $g$ as expected for the film contribution in a range typical of ferroelectric materials.[11] Therefore, R1-CPE1 describes the sample-electrode and/or sample-substrate interface, and R2-CPE2 the BFO film. In polycrystalline BFO films, a thickness dependency of the film dielectric constant was reported,[51] which was not the case in our epitaxial films.

BMO films did show a thickness dependence of C2 by normalizing by $g$, but identification of interface and film was still achieved unequivocally. Spectra were cut off at high frequency (~ 2 MHz), where data proved to be unreliable due to irregular $z'$ behavior.

### B. Temperature dependent analysis

We carried out temperature dependent analysis on coherently strained 50 nm BFO and BMO films using both equivalent circuit models containing (a) R-CPEs and (b) Voigt elements, and the fitted parameters were plotted vs $T$. Representative fits for BFO at 200ºC are demonstrated in the –$Z''$ vs $Z'$ plots in Fig. 4 (a,b,c). In the frequency range where the film contribution R2-C2 is dominant, both models resulted in a reasonable fit. The spectrum in Fig. 4(a) is dominated by the interface contribution. Fig. 4(b) presents the intermediate frequency regime and allows identification of the film contribution, Fig. 4(c) shows the residual resistance R0 on the real axis and a change of sign in $Z''$, indicating the presence of the inductive component L0 at high frequency. BMO film spectra showed smaller differences between interface and film contribution and overlap of two semicircles of similar dimension. The film capacitance values (C2) are shown in Fig. 5, obtained from the Voigt model in F cm$^{-1}$ and from the R-CPE circuit in corrected units of F cm$^{-1}$. It can be seen that the capacitance values are always comparable for both circuits, which justifies the use of R-CPE elements to obtain dielectric properties. We obtained values of 320 ± 75 (BFO) and 450± 100 (BMO) for the real part of the film dielectric constant $\varepsilon'_2$; data points at high and low temperatures were omitted from the analysis:

In the full temperature range 25ºC - 300ºC film values from R2-CPE2 and (R2-C2)$_{Voigt}$ for BFO and BMO showed considerable error due to strong overlap with the low frequency R1-CPE1 / (R1-C1)$_{Voigt}$ contribution (Fig. 4(b)) and with the high frequency contributions R0 and L0 (Fig. 4(c)). From fits to the data it was clear that the deviation of C2 vs $T$ from constant behavior (Fig. (5)) at high temperatures is caused by an increased overlap of residual resistance R0 and film contributions such that resolution of R2-C2 is less than ideal. The model was over-determined in this case, which was indicated by low values of $\chi^2$ < 0.5 [for a definition of $\chi^2$ see Ref. 52], implying that the fitted parameters can not be trusted. Contrarily, C2 deviations from constant behavior at low temperatures may be real effects, because $\chi^2$ ~1 indicated a valid fit. Such transitional C2 vs $T$ behavior at $T$ < 100ºC has been observed previously and may possibly be associated with effects from adsorbed water or moisture on the films.[53] The error bars displayed in Fig. 5 have been determined from R-CPE fits using commercial software (Z-View) and are believed to be also a good estimate for the errors for the fits obtained using our MatLab® software.

Both, BFO and BMO film dielectric constants $\varepsilon'_2$ are considerably larger than values reported previously, which were all recorded using limited range frequency measurements. In the Voigt model, the mean time constants $\bar{\tau}$ of the film contribution showed consistent standard deviations $\Delta\tau$ between 41% and 47% (BFO) and between 82% and 84% (BMO) of the respective $\bar{\tau}$ value. The standard deviation may be indicative of the degree of disorder in the material. The $\rho$ vs $1/T$ curves of the film contributions are shown in Fig. 6, indicating thermally activated charge transport with activation energies of 0.60 eV ± 0.05 eV (BFO) and 0.25 eV ± 0.03 eV (BMO). The BFO values are in a range suggesting that charge transport is dominated by oxygen vacancies,[54,55] whereas the lower BMO value may indicate an electronic contribution. The interface resistances R1 were two orders of magnitude (BFO) and by a factor of ~ 2 (BMO) higher than the respective film contributions (see Fig. 4), which suggests a blocking effect of the electrodes, as commonly observed at metal-ferroelectric interfaces.[49] This major finding of our work implies that the high resistivity values reported previously for multiferroic films may be strongly affected by electrode interface effects, and film resistivity and leakage behavior in these studies may have been misinterpreted. The above analysis was repeated for spectra collected at an applied magnetic field of 0.5 T, but no magneto-capacitance effects were found for BFO and BMO films at room temperature. It has to be noted though that magnetocapacitance effects are expected to be largest near the magnetic phase transitions, and further investigations at higher fields and low temperature are required. In the 50 nm BFO layer the interface resistance R1 showed an activation energy of ~ 0.6 eV and ~ 0.31 eV in the BMO layer, both in a similar range as the thin film values. Interpretation of the interface R1-CPE1 is not possible, because the origin of the barrier is unclear due to the different types of interface being present, namely Pt/BFO and Nb-STO/BFO.

Likewise, interpretation of the temperature dependence of R0 is not meaningful, because the resistance contains contributions from the metallic Pt electrodes and Cu measurement leads, and the Nb-STO substrate. The nominal resistance R0 was 4.5 Ω - 8 Ω between 50ºC - 300ºC for the BFO samples and 4 Ω - 6.3 Ω in the BMO samples. Both showed a positive temperature coefficient of resistance. Deconvolution of the different contributions was not feasible.

## VI. CONCLUSIONS

We conclude that IS is a technique that enables full characterization of the dielectric and resistive properties of multiferroic epitaxial thin films. IS has to be performed over a wide frequency and temperature range to reveal the composite character of multiferroic thin film sample response. Numerical equivalent circuit fitting was required in order to obtain reliable values for the intrinsic dielectric constant and resistivity of BFO and BMO epitaxial layers. A Voigt element has been used to provide a physically meaningful way to describe non-Debye behavior, which we propose as a possible replacement for phenomenological CPE circuits. From this study it is clear that limited frequency range measurements on ferroelectric thin films can be dominated by the interface response and may be inappropriate to extract reliable information.

## ACKNOWLEDGMENTS

We are grateful to M.E. Vickers for help with the x-ray analysis, and J.F. Scott and N.D. Mathur for useful discussions. This work was funded by the Royal Society (F.M.), an EU Marie Curie Fellowship (W.E.) and the Leverhulme Trust (R.S.). R. S. would like to thank J. Aguilar for the kind invitation for a research visit to FIME at UANL Monterrey (Mexico).

**References**

[1] N.A. Spaldin and M. Fiebig, *Science* **309**, 391 (2005).

[2] W. Eerenstein, N.D. Mathur, and J.F. Scott, *Nature* **442**, 759 (2006).

[3] S.-W. Cheong and M. Mostovoy, *Nature Materials* **6**, 13 (2007).

[4] R. Ramesh and N.A. Spaldin, *Nature Materials* **6**, 21 (2007).

[5] J.R. Teague, R. Gerson, and W.J. James, *Solid State Commun*. **8**, 1073 (1970).

[6] H. Chiba, T. Atou, and Y. Syono, *J. Solid State Chem.* **132**, 139 (1997).

[7] J. Wang, J. Neaton, H. Zheng, V. Nagarajan, S.B. Ogale, B. Liu, D. Viehland, V. Vaithyanathan, D.G. Schlom, U. Waghmare, N.A. Spaldin, K.M. Rabe, M. Wuttig, and R. Ramesh, *Science* **299**, 1719 (2003).

[8] T. Zhao, A. Scholl, F. Zavaliche, K. Lee, M. Barry, A. Doran, M.P. Cruz, Y.H. Chu, C. Ederer, N.A. Spaldin, R.R. Das, D.M. Kim, S.H. Baek, C.B. Eom, and R. Ramesh, *Nature Materials* **5**, 823 (2006)

[9] T. Kimura, S. Kawamoto, I. Yamada, M. Azuma, M. Takano, and Y. Tokura, *Phys. Rev. B* **67**, 180401(R) (2003).

[10] J.R. Macdonald, *Impedance spectroscopy*, John Wiley & Sons, New York, 1987.

[11] J.T.S. Irvine, D.C. Sinclair, and A.R. West, *Adv. Mater*. **2**, 132 (1990) 132.

[12] S. Song and F. Placido, *J. Statist. Mech.: Theory Exp.* **OCT**, P10018 (2004).

[13] V.V. Sazonov, *Normal approximation: some recent advances*, Springer, Berlin 1981

[14] S. Rodewald, J. Fleig, and J. Maier, *J. Am. Ceram. Soc*. **84**, 521 (2001).

[15] S.H. Liu, *Phys. Rev. Lett*. **55**, 529 (1985).

[16] B.H. Armstrong, *J. Quant. Spectrosc. Ra*. **7**, 61 (1967).

[17] V.R. Palkar, J. John, and R. Pinto, *Appl. Phys. Lett*. **80**, 1628 (2002).

[18] Y.P. Wang, L. Zhou, M.F. Zhang, X.Y. Chen, J.-M. Liu, and Z.G. Liu, *Appl. Phys. Lett.* **84**, 1731 (2004).

[19] K.Y. Yun, D. Ricinschi, T. Kanashima, M. Noda, and M. Okuyama, *Jpn. J. Appl. Phys*. **43**, L647 (2004).

[20] J. Dho, X. Qi, H. Kim, J.L. MacManus-Driscoll, and M.G. Blamire, *Adv. Mater*. **18**, 1445 (2006).

[21] Y.-H. Chu, Q. Zhan, L.W. Martin, M.P. Cruz, P.-L. Yang, G.W. Pabst, F. Zavaliche, S.-Y. Yang, J.-X. Zhang, L.-Q. Chen, D.G. Schlom, I.-N. Lin, T.-B. Wu, and R. Ramesh, *Adv. Mater*. **18**, 2307 (2006).

[22] C. Michel, J.-M. Moreau, G.D. Achenbach, R. Gerson, and W.J. James, *Solid State Commun*. **7**, 701 (1969).

[23] F. Kubel and H. Schmid, *Acta Cryst*. B **46**, 698 (1990).

[24] G.A. Smolenskii, V. Yudin, E.S. Sher, and Y.E. Stolypin, *Sov. Phys. JETP* **16**, 622 (1963).

[25] S.V. Kiselev, R.P. Ozerov, and G.S. Zhdanov, *Sov. Phys. Dokl*. **7**, 742 (1963).

[26] I. Sosnovska, T. Peterlin-Neumaier, and E. Steichele, *J. Phys. C Solid State Phys*. **15**, 4835 (1982).

[27] W. Eerenstein, F.D. Morrison, J. Dho, M.G. Blamire, J.F. Scott, and N.D. Mathur, *Science* **307**, 1203a (2005).

[28] H. Béa, M. Bibes, A. Barthélémy, K. Bouzehouane, E. Jacquet, A. Khodan, J.-P. Contour, S. Fusil, F. Wyczisk, A. Forget, D. Lebeugle, D. Colson, and M. Viret, *Appl. Phys. Lett*. **87**, 072508 (2005).

[29] M. Bai, J.L. Wang, M. Wuttig, J.F. Li, N.G. Wang, A.P. Pyatakov, A.K. Zvezdin, L.E. Cross, and D. Viehland, *Appl. Phys. Lett*. **86**, 032511 (2005).

[30] J. Wang, A. Scholl, H. Zheng, S.B. Ogale, D. Viehland, D.G. Schlom, N.A. Spaldin, K.M. Rabe, M. Wuttig, L. Mohaddes, J. Neaton, U. Waghmare, T. Zhao, and R. Ramesh, *Science* **307** 1203b (2005).

[31] I.G. Ismailzade, *Phys. Status Solidi B* **46**, K39 (1971).

[32] K.W. Yun, M. Noda, and M. Okuyama, *Appl. Phys. Lett*. **83**, 3981 (2003).

[33] Y.-H. Lee, J.-M. Wu, Y.-L. Chueh, and L.-J. Chou, *Appl. Phys. Lett*. **87**, 172901 (2005).

[34] K.Y. Yun, M. Noda, M. Okuyama, H. Saeki, H. Tabata, and K. Saito, *J. Appl. Phys*. **96**, 3399 (2004).

[35] X. Qi, J. Dho, R. Tomov, M.G. Blamire, and J.L. MacManus-Driscoll, *Appl. Phys. Lett*. **86**, 62903 (2005).

[36] E. Ascher, H. Rieder, H. Schmid, and H. Stoessel, *J. Appl. Phys.* **37**, 1404 (1966).

[37] T. Kimura, G. Lawes, T. Goto, Y. Tokura, and A.P. Ramirez, *Phys. Rev. B* **71**, 224425 (2005).

[38] N. Hur, S. Park, P.A. Sharma, J.S. Ahn, S. Guha, and S.W. Cheong, *Nature* **429**, 392 (2004).

[39] Y.Y. Tomashpolskii, Y.N. Venevtsev, and G.S. Zhdanov, *J. Exp. Theor. Phys*. **46**, 1921 (1964).

[40] M.M. Kumar, V.R. Palkar, K. Srinivas, and S.V. Suryanarayana, *Appl. Phys. Lett*. **76**, 2764 (2000).

[41] F. Sugawara, S. Iiida, Y. Syono, and S.-i. Akimoto, *J. Phys. Soc. Jpn.* **25**, 1553 (1968)

[42] H. Faqir, H. Chiba, M. Kikuchi, Y. Syono, M. Mansori, P. Satre, and A. Sebaoun, *J. Solid State Chem.* **142**, 113 (1999).

[43] T. Atou, H. Chiba, K. Ohoyama, Y. Yamaguchi, and Y. Syono, *J. Solid State Chem.* **145**, 639 (1999).

[44] W. Eerenstein, F.D. Morrison, J.F. Scott, and N.D. Mathur, *Appl. Phys. Lett.* **87**, 101906 (2005).

[45] R. Seshadri and N.A. Hill, *Chem. Mater.* **13**, 2892 (2001).

[46] A. Moreira dos Santos, S. Parashar, A.R. Raju, Y.S. Zhao, A.K. Cheetham, and C.N.R. Rao, *Solid State Commun.* **122**, 49 (2002).

[47] C.-H. Yang, T.Y. Koo, S.-H. Lee, C. Song, K.-B. Lee, and Y.H. Jeong, *Europhys. Lett.* **74**, 348 (2006).

[48] T. Zhao, H. Lu, F. Chen, S. Dai, G. Yang, and Z. Chen, *J. Cryst. Growth* **212**, 451 (2000).

[49] M. Dawber, K.M. Rabe, and J.F. Scott, *Rev. Mod. Phys.* **77**, 1083 (2005).

[50] C.H. Hsu and F. Mansfeld, *Corrosion* **57**, 747 (2001).

[51] S. Yakovlev, J. Zekonyte, C.-H. Spolterbeck, and M. Es-Souni, *Thin Solid Films* **493**, 24 (2005).

[52] J.R. Taylor, *An Introduction to Error Analysis*, University Science Books, Mill Valley, 1982.

[53] R. Schmidt and A.W. Brinkman, *Studies of the temperature and frequency dependent impedance of an electroceramic functional oxide thermistor*, Adv.Funct.Mat. **accepted for publication**

[54] S. Zafar, R.E. Jones, B. Jiang, B. White, P. Chu, D. Taylor, and S. Gillespie, *Appl. Phys. Lett.* **73**, 175 (1998).

[55] K. Nomura and S. Tanase, *Solid State Ionics* **98**, 229 (1997).

**Figure Captions**

FIG. 1. (Color online) Impedance response of circuit elements on a phasor diagram: applied voltage $U(\omega t)$, current response $I_R$ for an ideal resistor in phase with $U(\omega t)$, $I_C$ for an ideal capacitor with a $-\pi/2$ phase shift, $I_{RC}$ for a series resistor-capacitor combination with phase shift $\delta$, $I_L$ for an inductor with phase shift $+\pi/2$, and $I_{CPE}$ for a CPE describing a non-ideal capacitor with a frequency independent phase shift $\gamma$ with respect to the ideal capacitor; one phase of the applied voltage corresponds to a $2\pi$ rotation of the $U(\omega t)$ arrow.

FIG. 2. (Color online) **(a)** Sample geometry for A.C. impedance measurements of insulating multiferroic layers on a low resistivity Nb-STO substrate using top-top Pt electrodes **(b)** Equivalent circuit model.

FIG. 3. (Color online) Capacitance values for C1 (◊,♦) and C2 (□, ■) for BFO films normalized to contact area $A$ (◊, □) and to the geometrical factor $g$ (♦,■) vs film thickness, data taken at 30ºC.

FIG. 4. **(a,b,c)** (Color online) $-Z''$-$Z'$ plots of a 50 nm thickness BFO thin film spectrum at 200ºC; □ = data; ▫ = Voigt model; ■ = CPE model. All graphs are $-Z''$ vs $Z'$ plots [Ω vs Ω].

FIG. 5. (Color online) **(a)** Plots of capacitance C2 from CPE fits corrected to F cm$^{-1}$ (♦,■) and from Voigt fits (◊,□) in F cm$^{-1}$ vs $T$; data taken from 50 nm BFO (□,■) and 50 nm BMO (◊,♦) films, black solid lines indicate where $\varepsilon'_2$ and $\rho_{dc}$ values were extracted, dashed lines deviations from approximately constant behavior **(b)** real part of the film dielectric constants $\varepsilon'_2$ (dimensionless) for BFO and BMO

**FIG. 6.** (Color online) Plots of R2 for 50 nm BFO (□,■) and BMO (◊,♦) films in Ω cm vs $1/T$ from Voigt (□,◊) and CPE (■,♦) fits; the activation energies were determined from the slopes of ln(R2) vs $1/T$ plots.

**FIG. 1.** (Color online)

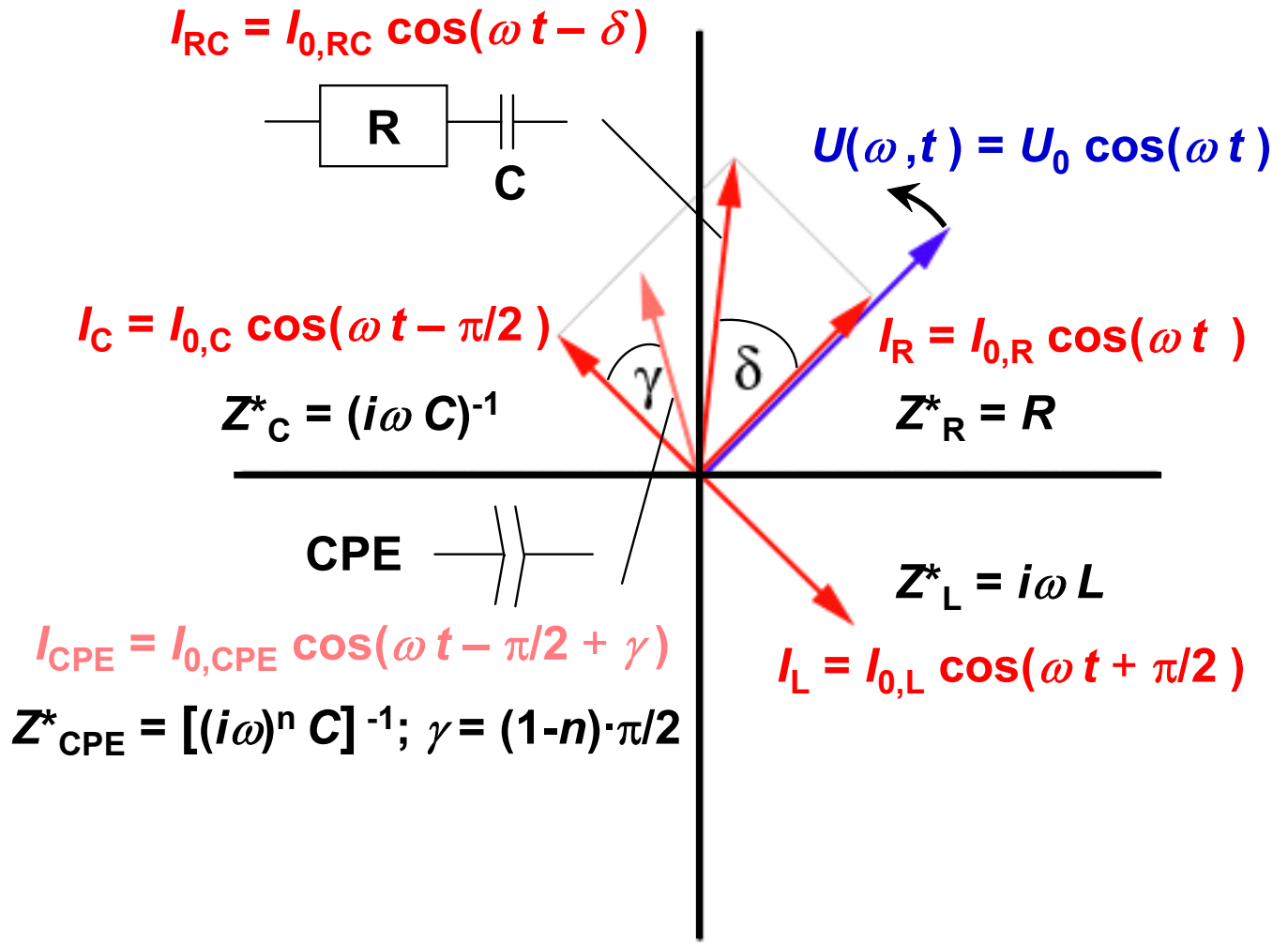

**FIG. 2.** (Color online)

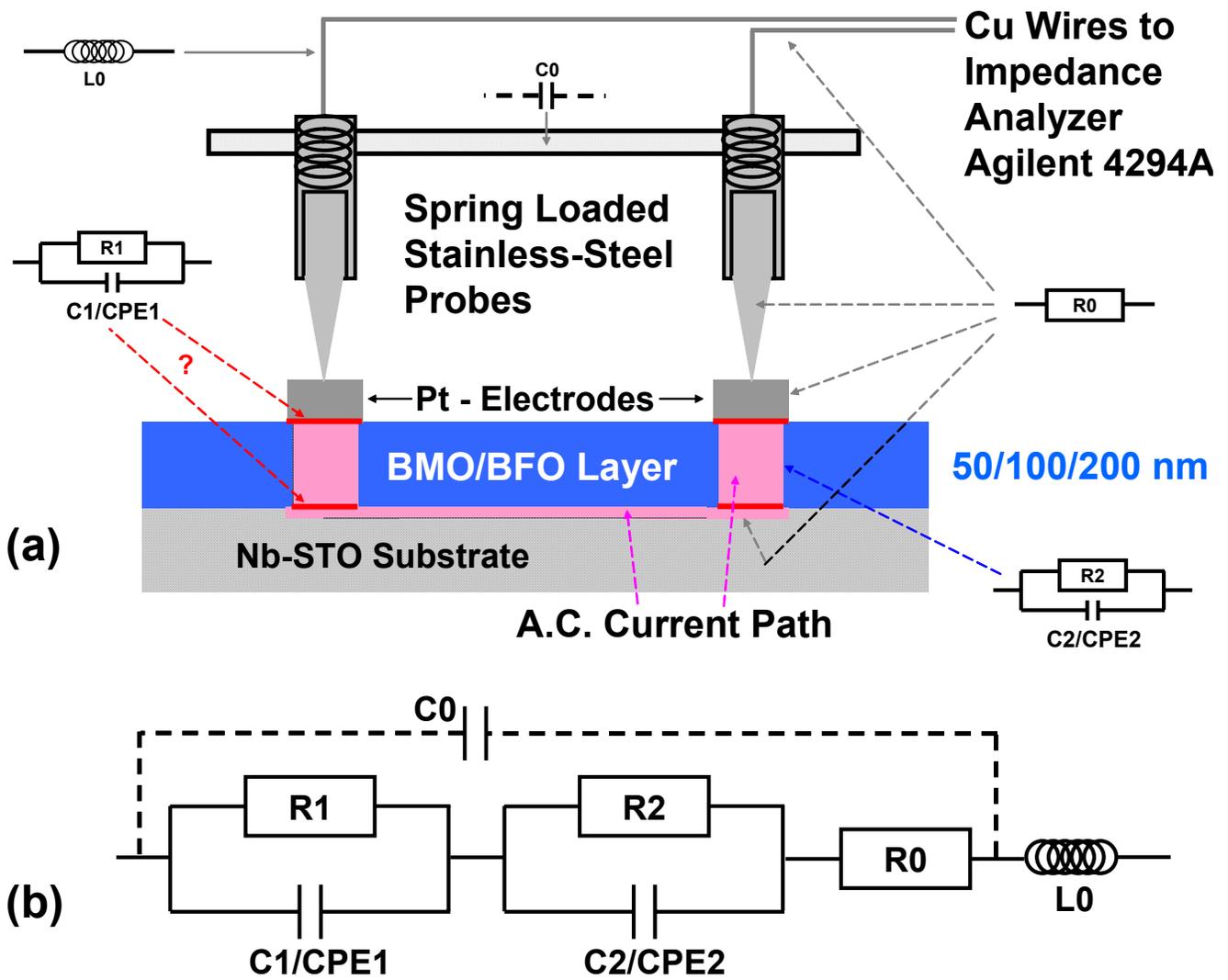

**FIG. 3.** (Color online)

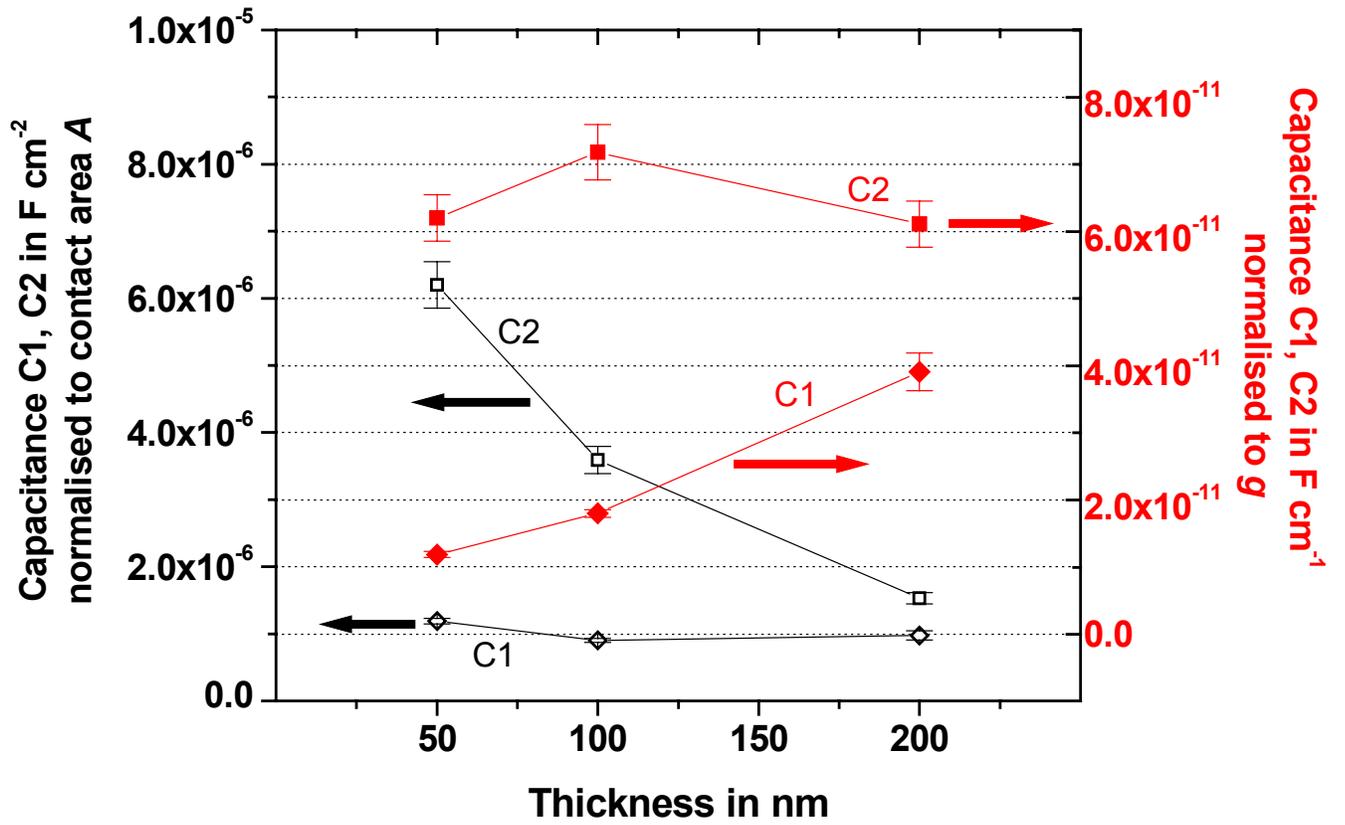

**FIG. 4.** (Color online)

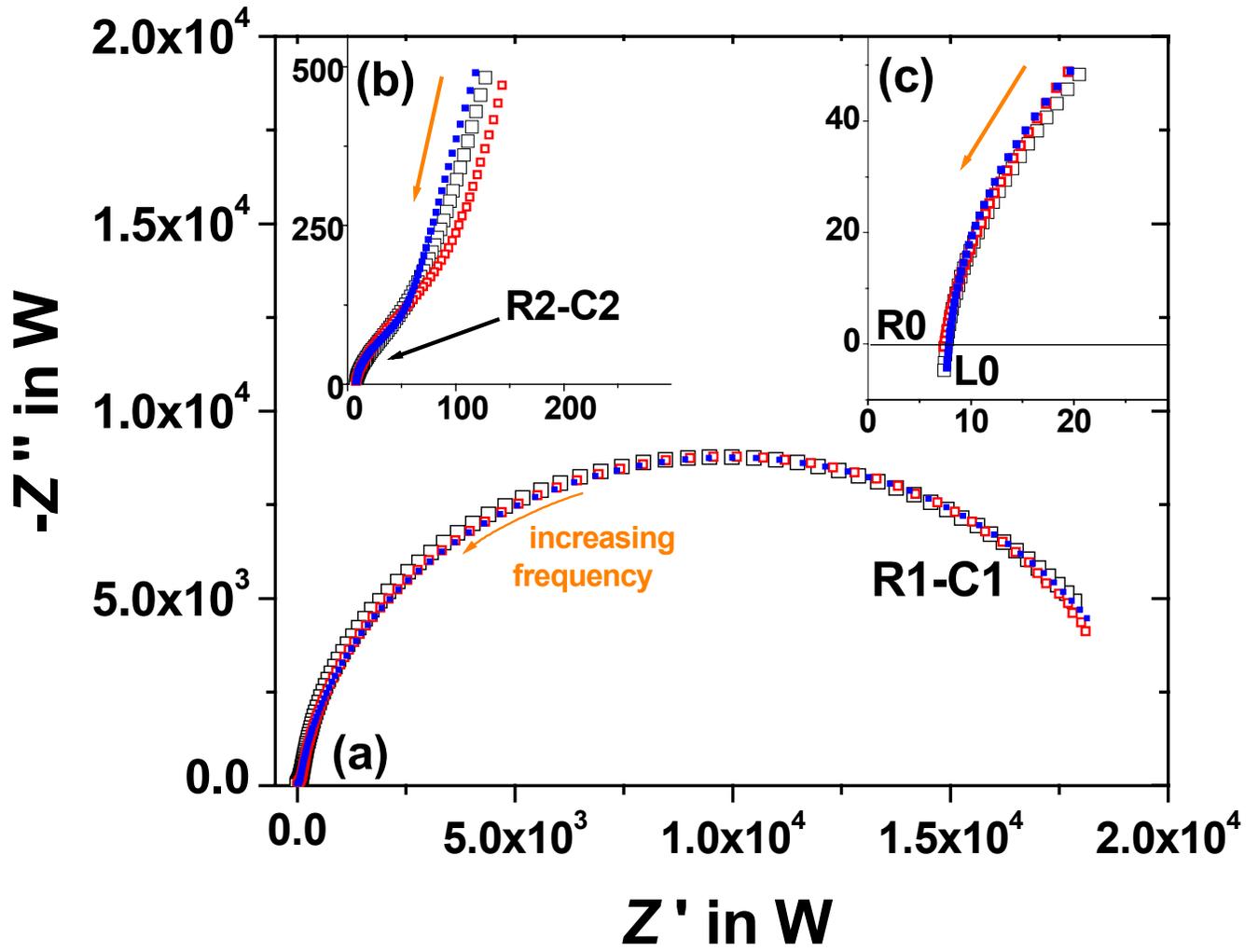

**FIG. 5.** (Color online)

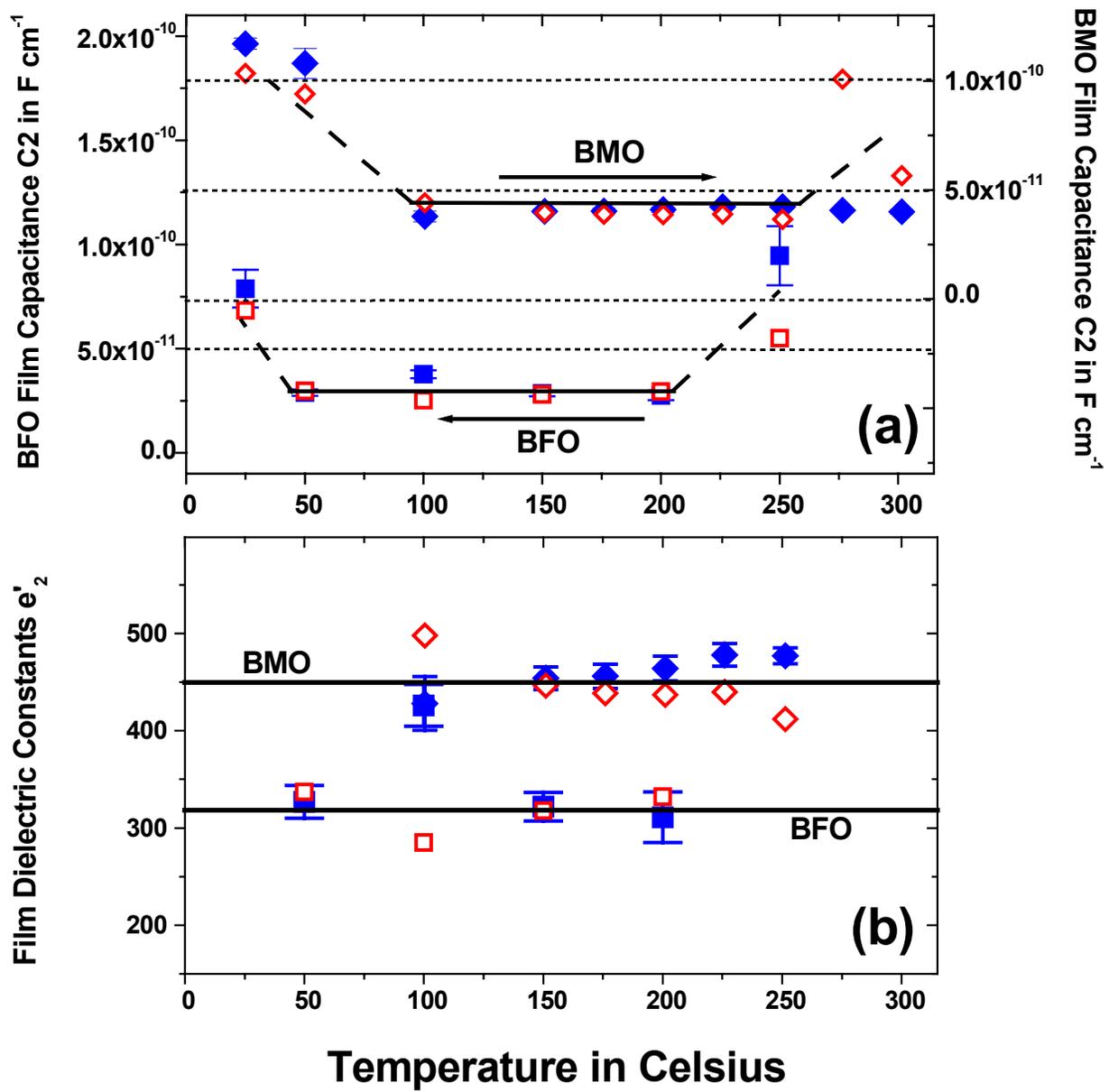

**FIG. 6.** (Color online)

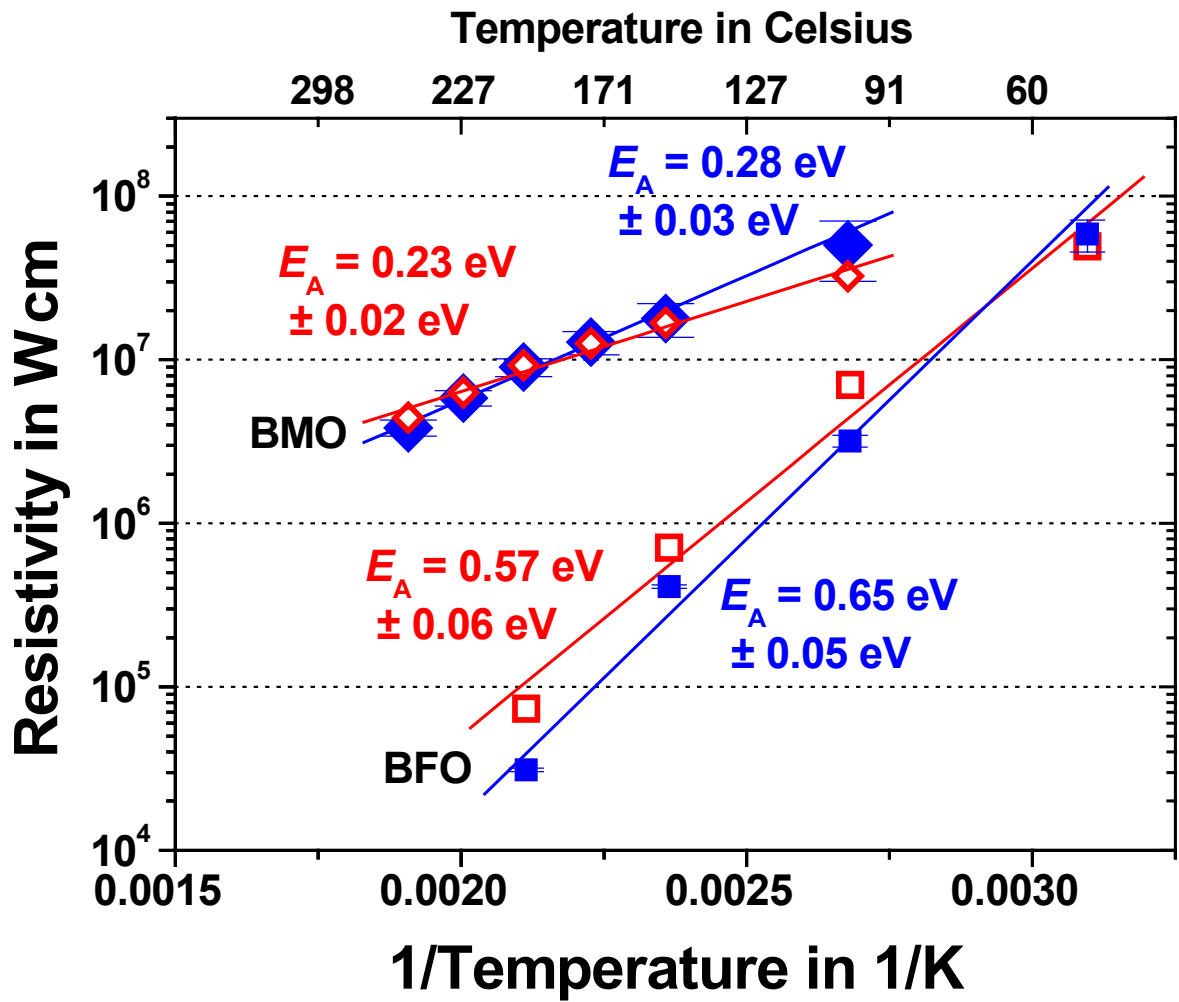